%% file: GCpsrSKA2024.tex
\documentclass[a4paper,11pt]{article}
\usepackage{aaskaiid}
\usepackage{orcidlink}
\usepackage{multirow}

\title{Pulsars in Globular Clusters With the SKAO}
\ShortTitle{Pulsars in GCs}

\author[1, 2]{M. Bagchi \orcidlink{0000-0001-8640-8186}}
\emailAdd{manjari@imsc.res.in}
\ShortName{Bagchi et al.} 

\author[3,4]{F. Abbate \orcidlink{0000-0002-9791-7661}}
\emailAdd{federico.abbate@inaf.it}

\author[4]{V. Balakrishnan \orcidlink{0000-0003-3244-2711}}
\emailAdd{vishnu@mpifr-bonn.mpg.de}

\author[3,4]{M. C. i Bernadich \orcidlink{0000-0002-2965-5911}}
\emailAdd{miquel.colomibernadich@inaf.it}

\author[5]{B. Bhattacharyya \orcidlink{0000-0002-6287-6900}}
\emailAdd{bhaswati@ncra.tifr.res.in}

\author[4]{A. Dutta \orcidlink{0000-0002-8950-5659}}
\emailAdd{adutta@mpifr-bonn.mpg.de}

\author[4]{P. C. C. Freire \orcidlink{0000-0003-1307-9435}}
\emailAdd{pfreire@mpifr-bonn.mpg.de}

\author[6]{K. Halley \orcidlink{0000-0001-8432-5889}}
\emailAdd{keh00032@mix.wvu.edu}

\author[7, 8]{J. W. T. Hessels \orcidlink{0000-0003-2317-1446}}
\emailAdd{j.w.t.hessels@uva.nl}

\author[5]{S. Kumari \orcidlink{0000-0002-3764-9204}}
\emailAdd{skumari@ncra.tifr.res.in}

\author[6]{D. R. Lorimer \orcidlink{0000-0003-1301-966X}}
\emailAdd{Duncan.Lorimer@mail.wvu.edu}

\author[3]{A. Possenti \orcidlink{0000-0001-5902-3731}}
\emailAdd{andrea.possenti@inaf.it}

\author[3, 9]{R. Nag \orcidlink{0009-0005-6754-0655}}
\emailAdd{rouhin.nag@inaf.it}

\author[10]{S. M. Ransom \orcidlink{0000-0001-5799-9714}}
\emailAdd{sransom@nrao.edu}

\author[11]{A. Ridolfi \orcidlink{0000-0001-6762-2638}}
\emailAdd{alessandro.ridolfi@uni-bielefeld.de}

\author[4]{V. Venkatraman Krishnan \orcidlink{0000-0001-9518-9819}}
\emailAdd{vkrishnan@mpifr-bonn.mpg.de}

\author[12, 13]{W. W. Zhu \orcidlink{0000-0002-6996-4364}}
\emailAdd{zhuww@nao.cas.cn}

\author[]{The SKA Pulsar Science Working Group}

\affiliation[1]{The Institute of Mathematical Sciences, Taramani, Chennai 600113, India}
\affiliation[2]{Homi Bhabha National Institute, Training School Complex, Anushakti Nagar, Mumbai 400094, India}
\affiliation[3]{INAF-Osservatorio Astronomico di Cagliari, Via della Scienza 5, I-09047 Selargius, Italy}
\affiliation[4]{Max-Planck-Institut f{\"u}r Radioastronomie, auf dem H{\"u}gel 69, 53121, Bonn, Germany}
\affiliation[5]{National Centre for Radio Astrophysics, Tata Institute of Fundamental Research, Ganeshkhind, Pune 411007, Maharashtra, India}

\affiliation[6]{Dept. of Physics and Astronomy, West Virginia University, Morgantown, WV 26506, USA}

\affiliation[7]{ASTRON, the Netherlands Institute for Radio Astronomy, Oude Hoogeveensedijk 4, 7990 PD, Dwingeloo, The Netherlands}

\affiliation[8]{Anton Pannekoek Institute for Astronomy, University of Amsterdam, Science Park 904, 1098 XH Amsterdam, The Netherlands }

\affiliation[9]{University of Cagliari, Monserrato University Campus - SP Monserrato-Sestu Km 0,700 - 09042 Monserrato (CA), ITALY}

\affiliation[10]{National Radio Astronomy Observatory, 520 Edgemont Road, Charlottesville, VA 22901, USA}

\affiliation[11]{Fakult\"at f\"ur Physik, Universit\"at Bielefeld, Postfach 100131, D-33501 Bielefeld, Germany}

\affiliation[12]{National Astronomical Observatories, Chinese Academy of Sciences, 20A Datun Road, Chaoyang District, Beijing 100101, China}

\affiliation[13]{Institute for Frontiers in Astronomy and Astrophysics, Beijing Normal University, Beijing 102206, China}

\abstract{Globular clusters (GCs) are highly efficient factories of radio pulsars: per unit of stellar mass, they contain about 1000 times more pulsars than in the Galactic field. Thus far, 345 radio pulsars have been found in GCs. These can be used as precision probes of the structure, gas content, magnetic field, and dynamic history of their host clusters; some of them are also highly interesting in their own right because they probe exotic stellar evolution scenarios, the physics of dense matter, accretion, gravity, etc. One of them (PSR~J0514$-$4002E) might even be the first pulsar - black hole system known. Deep searches with SKA telescopes will only require one to a few tied-array beams, and can be done during early commissioning of the telescopes, before an all-sky pulsar survey using hundreds to thousands of tied-array beams is feasible. Even a conservative approach predicts discoveries only with the core of SKA-MID AA*. Eventually, SKA-MID AA4 is expected to increase the number of discoveries even more, leading to more than doubling the current known population. Thus, a dedicated search for pulsars in GCs will fully utilise the best possible natural laboratories to study various branches of physics and astrophysics, including the properties of dense matter, stellar evolution, and the dynamical history of these GCs.}

\begin{document}

\maketitle

\section{Introduction: The Science of Globular Cluster Pulsars}

Globular clusters (GCs) are spherical, gravitationally bound clusters of stars containing from tens of thousands to millions of stars. They are especially known for their extremely high stellar densities, which in the cores of some GCs can reach $\sim 10^6 \, \rm M_{\odot} \, pc^{-3}$. There are more than 150 GCs associated with our Galaxy\citep{Harris1996}\footnote{The revision from December 2010 contains a list of 157 GCs, and is available at http://physwww.mcmaster.ca/$\sim$harris/mwgc.dat. On the other hand, a more recent compilation by Holger Baumgardt contains 165 GCs https://people.smp.uq.edu.au/HolgerBaumgardt/globular/parameter.html. Most of our simulations in this work is based on the Harris catalogue}; these are usually characterised by extremely old stellar populations, generally $> 10 \, \rm Gyr$.

Compared with the Galactic field, GCs contain thousands of times more Low-Mass X-ray binaries (LMXBs) per unit of stellar mass \citep{Clark1975}. This over-abundance is thought to be related to their extreme stellar densities, which create the conditions for a high rate of stellar collisions and interactions \citep{Sigurdsson1993, Sigurdsson1995}. In such collisions and exchange encounters, many neutron stars (NSs) that would otherwise be dead (of which there are about $10^9$ in our Galaxy) can find a new main-sequence companion. The evolution of main-sequence stars then leads, in tight binaries, to Roche lobe overflow, resulting in the NS accreting matter and angular momentum from its companion; thus forming the observed LMXBs.

The result of the evolution of LMXBs is in many cases a ``millisecond pulsar'' \citep[MSP,][]{Alpar1982,Radhakrishnan1982,tv2023}, a special type of radio pulsar that spins hundreds of times per second and has a much weaker magnetic field (and resulting spin-down rate) than most normal pulsars. As one might predict from the over-abundance of LMXBs in GCs, MSPs are similarly over-abundant \citep{Camilo2005, Ransom2008, Freire2013}. With the help of N-body simulations, \citet{ykc2019} found that average GCs contain up to 10 - 20 MSPs, while a very massive GC might contain close to 100 MSPs. Indeed, since 1987 \citep{Lyne1987}, searches of Galactic GCs have thus far discovered 345 pulsars in 45 clusters.

Pulsars in GCs are interesting, mainly because stellar interactions don't stop when MSPs form. Especially, in GCs where the rate of interactions per star ($\gamma$) is large \citep{Verbunt2014}, for any particular MSP, there is a large probability of further interactions in the lifetime of the cluster. Relatively close encounters increase the eccentricities of many of these binary pulsars. Indeed, many of the MSPs with Helium white dwarf (He-WD) companions have much larger eccentricities than those seen in the Galactic field. However, in some interactions, a random star in the GC can make such a close approach to the binary pulsar that chaotic interactions occur. In most cases, the MSPs exchange their low-mass He-WD companions for the more massive intruders, forming eccentric systems that are unlike any found in the Galactic disk \citep{Anderson1990,ps1991,Freire2004,Lynch2012,drk2015,ridolfi2021meerkatgc,ridolfi2022ngc1851disc,balakrishnan2023,prf2024}. In most of the cases, these new companions are massive WDs or NSs, however, in one case  (PSR~J0514$-$4002E, located in NGC~1851), the companion mass falls in the `low mass-gap' range, a range of mass, which, according to our current theoretical understanding augmented by observational hints, is too high for an NS to achieve and too low for a BH to exist. Confirming this would make this a `holy grail' type of system that would allow new tests of gravity theories \citep{Liu2014,FreireWex2024}.

In some of the GCs with the largest values of $\gamma$, especially in the core-collapsed GCs, many binaries get `ionized' \citep{br09a, br09b} too. For this reason, more than 95\% of MSPs in some of these clusters are isolated \citep{abbate2022ngc6624disc,arf2023,Corongiu2024,Yin2024,zwl2024,wpq2024,sdd2024}. More interestingly, some of the pulsars in these very dense GCs are much younger and far more energetic than what one would expect given the extreme ages of these stellar systems, having the largest amounts of $\gamma$-ray emission seen in MSPs \citep{Freire2011,Johnson2013,grf2022,zxw22}. The reasons for this are not yet entirely clear: this could be because some LMXBs get disrupted during their long accretion phase, leaving behind fast-spinning, but not yet fully recycled (i.e., high B-field) pulsars \citep{Verbunt2014}. However, it is possible that in some cases; new, highly magnetized and extremely powerful pulsars or magnetars are being produced in exotic ways, like WD mergers or accretion-induced collapse \citep[see e.g.][]{Boyles2011, kfp2023}. The existence of such unusual NSs is suggested by the discovery of fast radio bursts in a globular cluster in M81, which likely requires an extremely magnetized NS \citep{kmn22}. This improves the prospects for the discovery of much more extreme objects in Galactic GCs. This claim is supported by very recent simulations\citep[DRAGON-III]{wcs2025} where N-body simulations resulted in multiple interesting objects like pulsars, X-ray binaries, black hole-black hole mergers, intermediate mass black holes, etc even within first 100 Myr of life of a GC! 

Apart from their intrinsic interest, GC pulsars can even be used to probe properties of GCs, making these objects even more interesting. We discuss this aspect in detail in Section \ref{subsec:clusterproperties_usingpulsars}.

Population simulations, based on the results of deep pulsar searches in the past decade, show that the Galactic GCs harbour many more pulsars still to be discovered \citep{Bagchi2011, Chennamangalam2013, Turk2013, Yin2024}. In particular, once extrapolated to the total sample of 157 known GCs in the Milky Way, \citet{Turk2013} predicted a population range of potentially observable pulsars (i.e., those beamed towards us) between $600 - 3700$ (95\% confidence level). Furthermore, \citet{zh22} predicted 600$-$1500 X-ray detectable MSPs in GCs, and it is expected that most of them would be radio detectable too. These numbers suggest that the majority of the MSP population in GCs remains undiscovered, with up to an order of magnitude more pulsars awaiting discovery. The reason for this is simple - the large distances of GCs mean that, in most GCs, only the very brightest radio pulsars are found. This implies that pulsar surveys in GCs are fundamentally limited by sensitivity, and secondarily by processing techniques; current popular search techniques are still inadequate to find fast-spinning pulsars in compact, highly accelerated orbits. The above conclusion is supported by the recent radio imaging study of Terzan 5 by \citet{usc2026}. Finding more pulsars is always significant, as experience indicates that any significant growth in the known population inevitably leads to the discovery of exotic and rare new kinds of pulsar systems, like pulsar-BH binaries \citep{bdf24} or planets orbiting pulsar binaries \citep{Sigurdsson2003} or possibly triple star systems \citep{ransom2014} that can lead, for instance, to new tests of gravity theories \citep[e.g.][]{bt2014}.

The regions of interest around the centres of GCs, i.e., the regions within their half-light radii, are only a few arc minutes across. This is reasonably well matched to the fields of view of sensitive radio telescopes. This means that, unlike for other types of pulsar surveys (where the sky area being covered also matters), surveys for pulsars in GCs are mostly limited by the sensitivity of the radio telescopes used in the survey. Hence, sensitive telescopes like the SKA-MID and SKA-LOW will definitely increase the number of pulsars, especially pulsars that are intrinsically faint or pulsars that are located in distant GCs.

\begin{figure}[h]
    \centering
	\includegraphics[width=1.0\columnwidth]{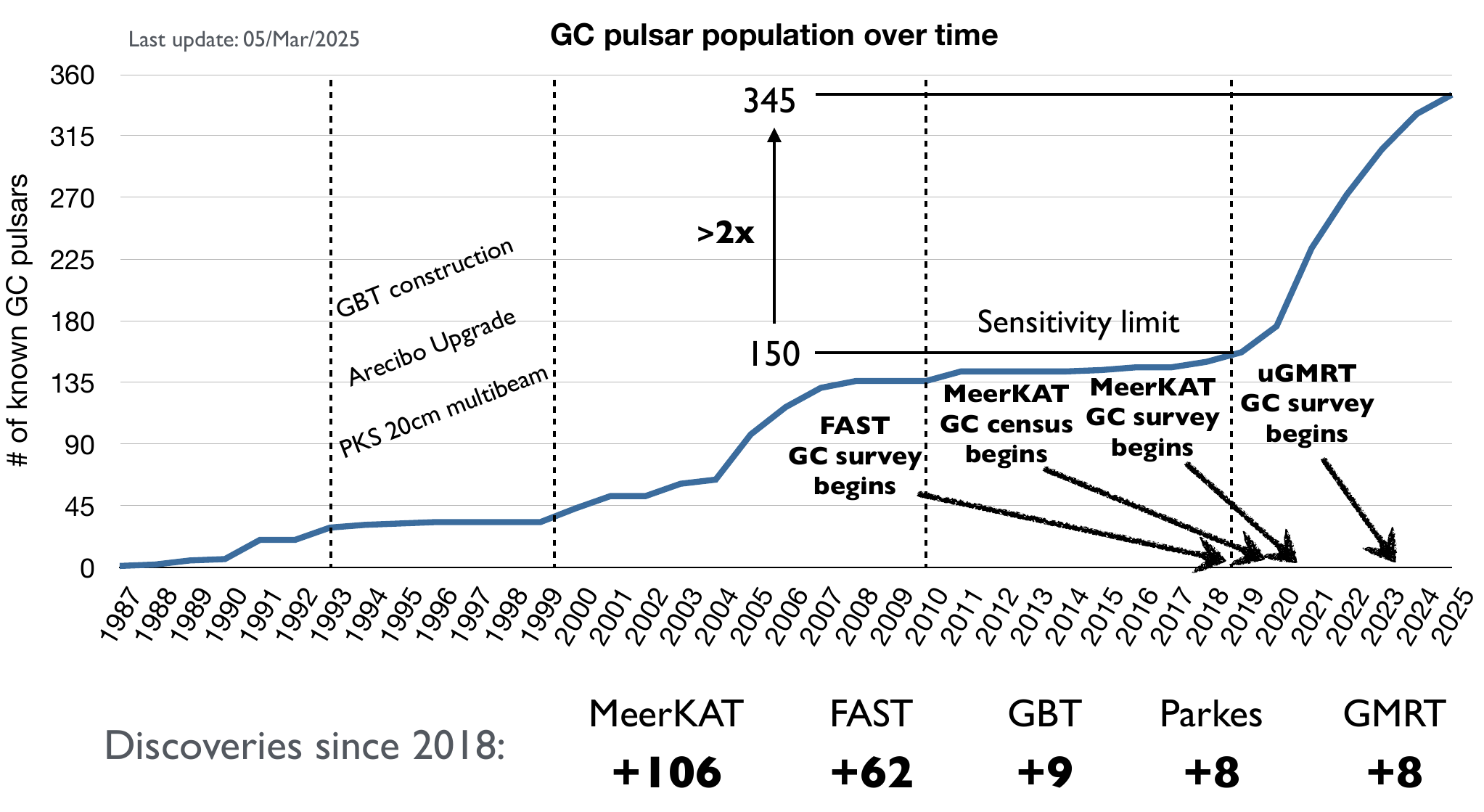}
    \caption{Pulsar population in GCs with time.}
    \label{fig:GC_PulsarpopulationOverTime}
\end{figure}

A consequence of this is the fact that, since the earliest handful of GC pulsar discoveries in the late 1980's / early 1990's \citep[e.g.][]{Lyne1987, Lyne1988, Manchester1990, Anderson1990, Manchester1991, Kulkarni1991, Manchester1991, Anderson1993}, the known population of GC pulsars has increased in a stepwise manner (see Fig.~\ref{fig:GC_PulsarpopulationOverTime}). The steps correspond to times when new, more sensitive telescopes and receivers became available, as described below:

1) The step in the 2000's is caused by the start of operations of the very sensitive Parkes multibeam receiver \citep[e.g.][]{Camilo2000,Possenti2003}, the start of operations of the Green Bank Radio Telescope (GBT) equipped with an exceptionally sensitive S-band receiver and a broadband pulsar back-end \citep[e.g.][]{Ransom2004, Ransom2005, Hessels2006, Freire2008a, lrf2011, Lynch2012}, and the resumption of operations of the upgraded Arecibo telescope, which included the sensitive Gregorian L-band receiver \citep{Hessels2007}. By the late 2000's, these surveys were completed and the discovery rate slowed down dramatically.

2) The larger, ongoing increase in the number of GC pulsars, which started in 2019, is mostly caused by the arrival of more sensitive radio telescopes, especially two SKAO related projects: the Five hundred meter Aperture Spherical Telescope (FAST) in the Northern hemisphere \citep{prl2020, pqm21fast,plj23,lpz2023,Yin2024, ypr2021, wps2020, wpq2024,zwl2024} and MeerKAT in the Southern hemisphere \citep[e.g.,][]{bailes2020meerkatfacility,ridolfi2021meerkatgc,ridolfi2022ngc1851disc,abbate2022ngc6624disc,chen2023omegacen,prf2024}. The upgraded GMRT \citep{gupta17ugmrt} is also contributing at lower frequencies, finding steep-spectrum pulsars \citep{grf2022,drf2024}, and the GBT is also contributing at higher frequencies, finding high dispersion measure (DM) pulsars \citep{sdd2024,mmr+24}.
This ongoing increasing phase has more than doubled the number of GC pulsars and the number of GCs with known pulsars: at the time of the SKAO science case a decade ago \citep{Hessels2015}, there were 144 pulsars in 28 GCs. On the other hand, at the time of writing the present chapter, there are 345 pulsars (an increase of 201) in 45 GCs (an increase in 17). 

Apart from the large collecting areas and sensitive receivers, the increased rate of discoveries also owes much to the increased bandwidths of receivers and pulsar data recorders and to the latter's improved time and frequency resolution. These are critical for identifying the fastest rotating pulsars \citep{Hessels2006}. These surveys also benefited from the improved computational capabilities and improved search algorithms \citep[e.g.,][etc]{ar18}, which are essential not only for handling much larger data rates of these surveys, but also for identifying the compact and highly accelerated pulsars that are more likely to be scientifically rewarding, as discussed below.

\begin{figure}[h]
    \centering
	\includegraphics[width=1.0\columnwidth]{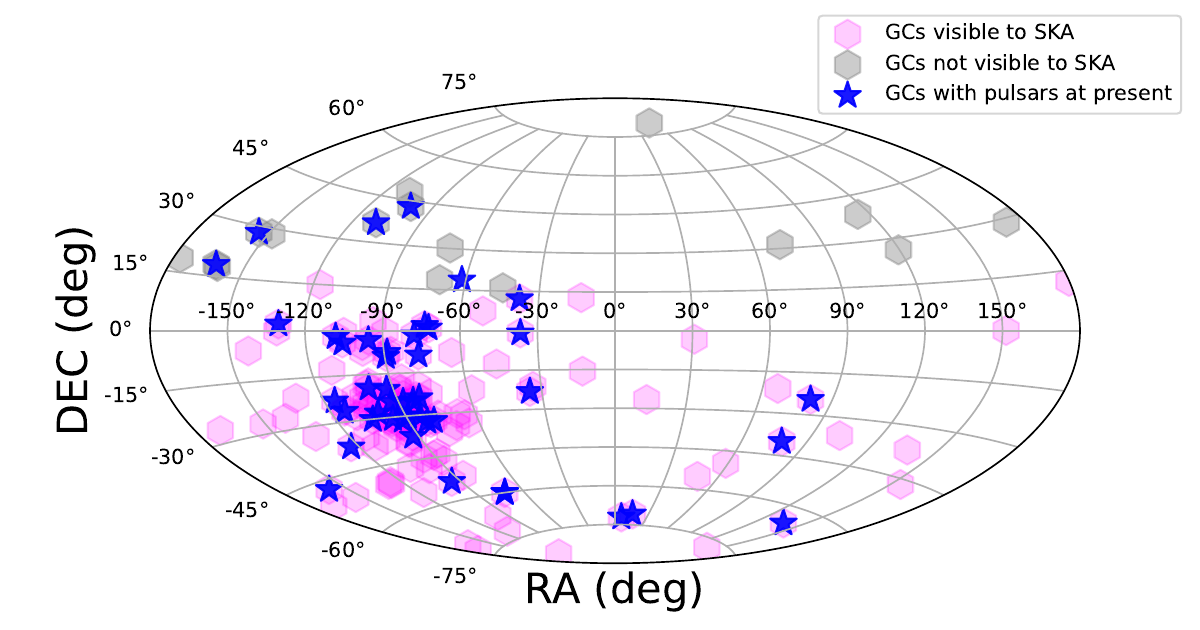}
    \caption{Sky distributions of Galactic globular clusters, demonstrating that there are many GCs in the sky (pink hexagons) observable by SKA telescopes and many of those GCs still do not have any pulsars discovered in them (pink hexagons without blue stars). GCs in the declination range $+15^{\circ}$ to $-90^{\circ}$ have been considered as visible by the SKA telescopes.}
    \label{fig:AllGC_withPulsars}
\end{figure}

Given the large number of undiscovered pulsars in GCs, and the increase in the rate of discoveries every time a new, more sensitive observing system comes online, it is clear that the increase in sensitivity in the Southern Hemisphere to be provided by the SKA telescopes will result in the discovery of many GC pulsars. This is supported by the fact that although FAST is more sensitive than MeerKAT, it has discovered fewer GC pulsars (62 vs 106). MeerKAT has a more advantageous position because more GCs are visible from the southern hemisphere (see Fig. \ref{fig:AllGC_withPulsars}). This advantageous position is crucially shared by the SKA telescopes. Furthermore, given the small areas of the sky to be searched, GCs can provide an early boom in the number of pulsars to be found by the SKA telescopes, as happened with MeerKAT.

The southern locations of most of the GCs  also mean that MeerKAT and the SKA telescopes will be able to search for new pulsars in many of these clusters - perhaps eventually in almost all Galactic GCs. This would be interesting mainly because it would raise the probability of finding unusual systems, but also because it would help provide more detailed information on the relation between the characteristics of the different pulsar populations and cluster properties.

\section{The science of GC pulsars}
\label{sec:science}

\subsection{Individual systems: }
\label{subsec:science_individual}

Individual systems that can be used for investigations in fundamental physics, like the Equation of State (EoS) of neutron-rich matter at super-nuclear densities and/or tests of gravity theories. Interesting individual systems can also provide novel insights for understanding stellar evolution and accretion physics \citep{Archibald2009, Papitto2013, Bassa2014, Stappers2014,Tauris2017,tv2023}. Although these science cases can also be addressed using Galactic pulsars, the unique environments of GCs favor the formation of exotic objects, the parameters of which may be even better suited for certain applications or constraining physical models.

Moreover, the study of some unique sub-classes of pulsars can be better done for GC pulsars, mostly due to their abundance. The examples include isolated MSPs, highly eccentric binaries, etc. These systems are also useful for studying stellar interactions, e.g., exchanges and close fly-bys, an example being NGC 1851A \citep{dfg2025}. For some sub-classes, a comparison between the GC populations and field populations might be interesting and are discussed below.

\begin{itemize}

    \item {\it Binary pulsars as test beds of gravity}: Radio pulsars have been used for tests of gravity theories (e.g., \citealt{agh2018,vcf2020,Kramer2021}, for a review see \citealt{FreireWex2024}). The unique characteristics of some GC pulsars open perspectives on new types of tests of gravity theories. For instance, as mentioned earlier, PSR~J0514$-$4002E, located in the GC NGC~1851 might have a black hole (low mass) companion \citep{bdf24}. This means that several theories of gravity that were previously untestable might now become testable for the first time. Additionally, MeerKAT timing might in the near future, do a test of the cosmic censorship hypothesis via the measurement of the Lense-Thirring effect caused by the rotation of the companion. Prospects of testing gravity with binary pulsars (regardless of whether those are in GCs or in the Galactic field) with the SKA telescopes have been discussed extensively in \citet{VenkatramanKrishnan1.2026.SKA}.

    Note that, at present, it is not possible distinguish between an black hole and a neutron star companion of PSR J0514$-$4002E using radio timing analysis with present day accuracy \citep{bdf24}, in the future continuous timing over a long time base line might improve the mass measurements of the masses of the system through improved estimates of Post-Keplerian parameters, especially Einstein delay parameter and the periastron precission. If the companion mass is found to be larger than $2.5~{\rm M_{\odot}}$, then it would hint at the possibility of being a black hole, as most of the neutron star equation of states fail to produce such higher masses. Additionally, a black hole in the mass gap region is likely to be formed bythe  mergers of two neutron stars, leading to a rapidly rotating black hole. In such a case, the black hole would give rise to the frame-dragging effect, which would impact both precession as well as orbital inclination \citep{ds1988}. However, this might be difficult to achieve unless the orbital inclination is measured accurately whcih would be possible if Shapiro delay and/or impact of light bending effect is measured \citep{dbb2023}. In summary, a continuous high-precision timing campaign might shed light to the true nature of the companion of PSR J0514$-$4002E and considering the scientific benefit, we would encourage such a timing campaign using SKA telescopes.

Although, at present, it is not possible to distinguish between a black hole and a neutron star companion of PSR J0514$-$4002E using radio timing analysis with current accuracy \citep{bdf24}, continuous timing over a longer time baseline in the future may improve the mass measurements of the system through better estimates of post-Keplerian parameters, especially the Einstein delay parameter and the periastron precession. If the companion mass is found to be larger than $2.5~{\rm M_{\odot}}$, this would hint at the possibility of it being a black hole, as most neutron star equations of state fail to produce such high masses.

Additionally, a black hole in the mass-gap region is likely to be formed through the mergers of two neutron stars, leading to a rapidly rotating black hole. In such a case, the black hole would give rise to the frame-dragging effect, which would impact both the periastron precession and the orbital inclination \citep{ds1988}. However, this may be difficult to detect unless the orbital inclination is measured accurately, which would be possible if the Shapiro delay and/or the effect of light bending is measured \citep{dbb2023}.

In summary, a continuous high-precision timing campaign may shed light on the true nature of the companion of PSR J0514$-$4002E, and considering the scientific benefits, we encourage such a timing campaign using SKA telescopes.

    \item {\it NS-NS (DNS) systems and gravitational waves astronomy}:  Presently, there are 35 pulsar binaries with the possibility of being DNS systems. Out of these, 24 systems are confirmed (based on considerations of stellar evolution) and a further 11 are unconfirmed candidates. Among these unconfirmed candidates, a total of 8 are in GCs\footnote{For an updated table of DNSs, see \url{https://www3.mpifr-bonn.mpg.de/staff/pfreire/NS_masses.html}.}. We expect many more DNS systems in Galactic GCs yet to be discovered.    
    Most of the latter have fast-spinning pulsars, e.g., Terzan 5ao, which has $P = 2.27$ ms \citep{prf2024}. This is one of the elements that suggests that these systems were formed via exchange encounters, as standard stellar evolution theories can not explain such rapidly rotating neutron stars in DNS binaries. Such exchange interactions even make it possible for the formation of double MSP binaries \citep{Sigurdsson1992}.

The large spin frequencies of the pulsars in DNSs make them potentially very interesting for gravitational wave astronomy \citep{phinney1991, kas2021, WBdM2022}. For the known DNSs in the Galactic disk, the adimensional spin parameter at the merging event is at most 0.032 \citep{Stovall2018}; this means that gravitational waveforms calculated for the mergers of circular and aligned spin axes generally do a good job recovering the observed signal, even though the spins in the Galactic DNSs are generally misaligned because of the second supernova in the system \citep{Tauris2017}. For the MSPs in globular cluster DNSs formed in exchange encounters, the dimensionless spin parameters would be one order of magnitude larger, inducing much larger spin signals in the gravitational waves coming from the merger; their spins are also not expected to be aligned with the orbital angular momentum. Furthermore, the constant interactions with other stars in the cluster might significantly increase the orbital eccentricity of the systems; without such interactions the eccentricity decreases monotonically. The detection of such eccentric, misaligned spin NS-NS mergers can be expected by the O4 run of LIGO-Virgo or the next generation ground-based gravitational wave detectors. Moreover, space-based gravitational wave detectors have the potential of detecting gravitational waves from NS-NS binaries at the early inspiral phase \citep{amaroseoane2023, mkkacv2025}.   
   
 \item {\it Ultra-Fast Rotators}: Unlike in the Galactic field, an MSP in a GC can, in principle, experience multiple episodes of recycling (due to dynamical encounters) and hence could be more effectively spun up to its limiting rotational period (as well as growing in mass). Perhaps, this is the reason for GCs harboring some of the fastest-spinning neutron stars known, including the 716-Hz record holder in Terzan 5 \citep{Hessels2006}. Though such sources are rare \citep{Hessels2007}, doubling or tripling the known population provides great prospects for pushing towards even faster rotation rates, perhaps even a sub-millisecond pulsar, which could strongly constrain the EoS \citep{Lattimer2001}, especially if the mass of the NS is measurable. 

 \item {\it Neutron Star Masses}: Many (at least 39) eccentric ($e > 0.01$) pulsar binary systems exist in GCs. For these, the mass of the NSs can be estimated with the help of periastron precession \citep[e.g.,][]{Freire2008a}. In the Galactic field, MSP orbits are often extremely circular ($e < 10^{-4}$), and prohibit such measurements in most cases. Besides the obvious implications of the highest-mass sources for constraining the neutron star EoS \citep{Antoniadis2013,Fonseca2021}, mapping the full MSP mass distribution is an important probe of their formation in supernovae, and their later `recycling' through accretion.  Once enough NS masses have been measured, the maximum NS mass should become apparent. 

\item {\it Radio pulsars to constrain EoS}: When binary pulsars with precise measurement of the NS mass is accompanied by the measurement of another EoS dependent parameter, it is possible to provide some constraints to dense matter EoS. One classic example is the constraint through measurement of the Lense-Thirring effect in the case of PSR~J0730-3039 A/B \citep{Kramer2021}.

\item {\it Young Pulsars:} As already mentioned, young pulsars in GCs \citep{Lyne1996, Boyles2011, prf2024, wpq2024}, provide interesting cases for exploring alternatives to the typical core-collapse supernova channel for forming neutron stars. 

\item {\it Spiders:} Spider pulsars are MSPs in compact orbits with low-mass companions. Many of spiders reside in GCs. The first GC spider Terzan 5A \citep{nttf90} was discovered within two years of the discovery of the first spider PSR  B1957+20 \citep{fst98}. The first MSP with a bloated main-sequence companion was also found in a GC \citep{DAmico2001}. Currently, the most compact binary pulsar known (having an orbital period of 53 minutes), PSR~J1953+1846E, is a spider system located in the GC M71 \citep{plj23}.

There are many theoretical works explaining the formation of spiders, e.g., \citet{ccth13}, with scope to improve the theories. Many observational phenomena, e.g., switch between radio MSP and low-mass X-ray binaries in quiescence and/or outburst of several sources \citep{Archibald2009, Papitto2013, Bassa2014, Stappers2014}, ultra-short period binary \citep{plj23}, are useful for this purpose. Hence, an increased population will be very good, especially to understand how the intricate evolutions of such systems differ in different environments existing in the Galactic field and in GCs.

Most of the spider pulsars undergo eclipses by the matter ablated from their companions, e.g., eclipses have been reported for about 20 out of 36 known GC spiders. These eclipses are often observed to be frequency-dependent, likely due to synchrotron absorption of radio emission and scattering processes by the ablated material and the magnetosphere of the companion  \citep{pebk88,tbep94,kb2021,Kudale2020,bhattacharyya2013,PolzinJ1810,kumari2024}. 
For some of these systems, the pulsar emission could partially pass the eclipsing material when the pulsar is in egress or ingress, which sometimes allows for indirect constraints \citep{ymc18,llm19,lbr23} or direct measurements of the electron densities and magnetic fields of the eclipsing materials or the pulsar wind \citep{csm20,mbz23,wwl23,kumaripol2024}. 

The wide bandwidth coverage of the SKA telescopes will play a major role in probing the frequency-dependent nature of such eclipses. High sensitivity studies by the SKA telescopes of the spider pulsars could map the variation of the rotation measure over the orbital phase and provide detailed insights into the properties and dynamics of the eclipsing medium, the interaction between the pulsar wind and companion outflows, and the magnetic field structures in these binaries. A good number of such binaries are in GCs and this number will increase even more by the GC pulsar surveys with the SKA telescopes.

Recent studies hint that the subclass of `black widows' with very small companion masses and orbital periods might even be a separate class known as `tidarrens' \citep{rgfz16, lkwthhl22}. Although all three presently known tidarrens are in the Galactic field, the possibility that these tidarrens were formed in GCs and then ejected out exists \citep{lkwthhl22}. An increase in the number of pulsars in GCs might lead to discoveries of more tidarrens. These systems will be useful in understanding the formation of isolated MSPs.

\item {\it Rotating Radio Transients:} `Rotating Radio Transients (RRATs)' are commonly thought to be either intermittent (long nulling) pulsars or giant pulse emitters whose non-giant pulses are too faint to be detectable by present radio telescopes \citep{zgd2007}. These are magnetospheric phenomena and should not depend on the environment. Hence, the existence of RRATs in GCs are expected. However, so far, all the known RRATs are in the Galactic field. This can be attributed mainly to the fact that, conventionally, GC pulsar surveys employ periodicity searches, and RRATs are discovered usually by single pulse searches, as well as the low number statistics, as there are only 336 RRATs known (https://rratalog.github.io/rratalog/) presently. 
Hence, regular employment of single pulse searches in GC pulsar surveys should reveal RRATs in GCs when the number of GC pulsars becomes nearly equal to the number of field pulsars.
Note that, conventionally, nulling pulsars or giant pulse emitters belong to the normal pulsar population. However, there are a handful of RRATs (5 out of 336) that are on the tail region of the normal pulsar population (spin period is less than 200 ms). Hence, increased population size of RRATs (be it in GCs or the field) will enable us to understand them as a population better. 

Moreover, RRATs in GCs might shed light to the alternative theories about their nature, e.g., infall of debris/asteroid through the magnetosphere model impacting the emission mechanism \citep{cs2008}. This model might prefer a dense environment of GCs (where asteroids are more likely). Even the non-detection of RRATs in GCs with SKA-telescopes (and regular single pulse searches) would be interesting (less likely) as it would compel us to think a completely different physical nature of RRATs.

As RRATs are discovered through single pulse search, the increased instantaneous sensitivity of SKA telescopes could lead to the first discoveries of RRATs in GCs.
As an example of the importance of the instantaneous sensitivity of a telescope in the studies of the single pulses from distant millisecond pulsars (as those embedded in GCs), we notice that the observations of MeerKAT collected about 1 giant pulse per second emitted from the millisecond pulsar PSR J1823$-$3021A in NGC6624 \citep{Abbate2020}, while only about 1 giant pulse per minute had been previously observed by less sensitive observations \citep{Knight2007}.

\end{itemize}

\subsection{Pulsar populations and their relation with cluster properties}
\label{subsec:population_clusterproperties}

\begin{figure}[h]
    \centering
    \includegraphics[width=0.48\columnwidth]{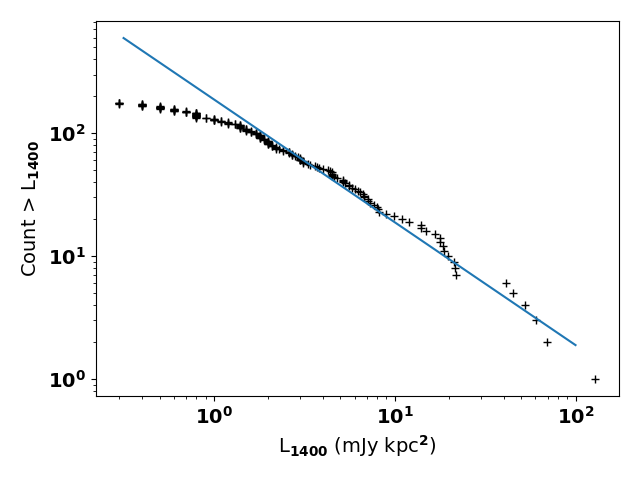} 
    \hfill 
    \includegraphics[width=0.48\columnwidth]{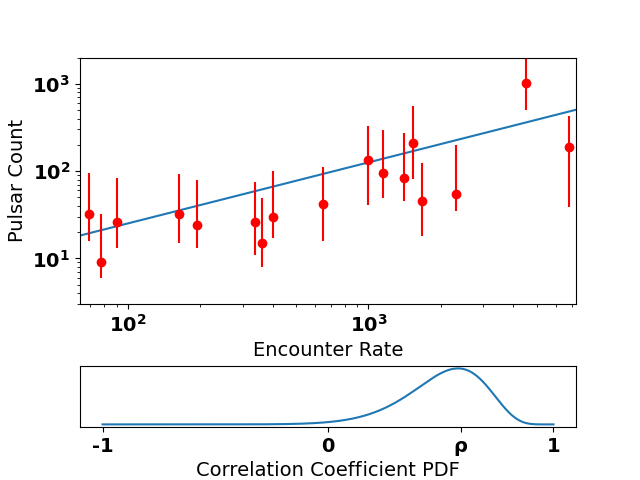} 
    \caption{Left: cumulative luminosity distribution of all 176 pulsars with published values of flux densities. The blue line is a power-law with an exponent of $\alpha = -1$. Right: the upper panel shows the correlation between estimated pulsar content, $N$, and the two-body encounter rate, $\Gamma$. The solid line compares these estimates to the best fit found by \citet{hct2010} in which $N = \Gamma^{0.7}$. The lower panel shows the significance of the correlation expressed as the probability density of Pearson's correlation coefficient.}
    \label{fig:lumfunc}
\end{figure}

While the sample of GC pulsars continues to grow, it is still significantly affected by the selection effect due to the limitation in observing sensitivity. Fig. \ref{fig:lumfunc} shows the impacts of this bias on the luminosity distribution which sees a decline in the number of fainter pulsars, which are the hardest to detect. A number of efforts to correct for this bias have been carried out to estimate the total pulsar content \citep[see, e.g.,][]{Bagchi2011,Chennamangalam2013}. It was found that an underlying log-normal luminosity distribution is the closest to the true distribution.

Scaling relationships between cluster properties have also been sought, with the most significant being the correlation between the pulsar population size and the two-body encounter rate \citep{hct2010,Turk2013}. Afterwards, a correlation between the pulsar content and the central escape velocity has also been reported \citep{Yin2024}. This is expected, as the central escape velocity and the two-body encounter rate are strongly correlated. In Fig.~\ref{fig:lumfunc}, we demonstrate the correlation between the pulsar content and the two-body encounter rate with the present data using the pulsar content for all 17 clusters in which at least two pulsars with measured flux densities are known. The pulsar content for each cluster was estimated assuming a log-normal pulsar luminosity function \citep{Bagchi2011} and the number of pulsars above the minimum luminosity for each cluster \citep{Chennamangalam2013}. With the help of N-body simulations, \citet{yefra23} presented a relation between the number of MSPs with various properties of GCs, e.g., the initial mass of the GC and its age. It would be interesting to see whether these results still hold when more MSPs are discovered in GCs. 

As mentioned earlier, the two-body encounter rate gives an estimate of the formation of binaries that are progenitors of present-day MSPs \citep{vf1987, Pooley2003}. However, once a binary is formed, it may undergo subsequent encounters. Hence, another important correlation has to do with the number of stellar encounters per binary, $\gamma$ \citep{Verbunt2014}. In some core-collapsed GCs, where this parameter is especially large, we observe a large over-abundance of isolated pulsars:  7 out of 7 in Terzan 1, 20 out of 21 in NGC 6517, 6 out of 6 in NGC 6522, 10 out of 12 in NGC 6624, 8 out of 9 in NGC 6752, 14 out of 15 in M15. In these GCs, we also observe many slow and high B-field pulsars.

\subsection{Using pulsars to understand cluster properties}
\label{subsec:clusterproperties_usingpulsars}

The cumulative population of pulsars can be used to probe the structure, proper motion, dynamical status, magnetic field, and gas content of the cluster itself \citep{Phinney1993, Meylan1997, Pooley2003}. Pulsar-determined quantities can constrain the still largely obscure GC evolution, as well as its relation with the history of the Galaxy.  For example, the pulsar-abundant cluster Terzan~5 contains at least two stellar populations of differing ages, and is possibly the pristine remnant of a building block of the Galactic bulge \citep{Ferraro2009}.  The same advances in computing that will allow the construction of the SKA telescopes will also allow advances in numerical simulations of GCs. Simulations of the stellar and dynamical evolution of stars in a GC will predict the observable population of neutron stars in each cluster, and this can be directly tested against observations. Neutron stars, as some of the heaviest objects in GCs, play a vital role by acting as an energy reservoir to counter the gravitational collapse of the GCs through the formation and disruption of binaries. The SKA telescopes will provide a census by looking for those neutron stars seen as pulsars, and their spin and binary properties will give important clues to the dynamical state of the cluster \citep[e.g.,][]{Verbunt2014, lyg2024}. 

\begin{itemize}

\item {\it Cluster Potential}: Acceleration and the first derivative of the acceleration (`jerk') in the cluster potential affect the pulsars' spin derivatives and orbital period derivatives, which can be used to probe the cluster potential in a very direct and unique way \citep{bra1987,Phinney1992}. In addition to helping determine various parameters related to the density profile, e.g., the core density, the core radius, etc. (as was done for 47~Tuc by \citealt{frk2017, Abbate2018} and Terzan 5 by \citealt{prf17}), modelling the cluster acceleration might also reveal the presence of an intermediate mass black hole (IMBH) at the core \citep{dcmp07}. In fact, over the past decades, there have been several efforts in this regard \citep{Abbate2018,Abbate2019b,Xie2024,Corongiu2024}. These studies have led to the detection of additional non-luminous mass in the centre that is not accounted for by the density profile of the clusters. This mass excess is thought to be caused either by a single IMBH or by a system of dark compact objects. These works have sparked new developments in the field of N-body simulations \citep{Gieles2018,Baumgardt2019,gkb2021,DcPB2021, fwt2024, Smith2024}. In order to obtain clear evidence of the presence or absence of IMBHs in GCs, more pulsars need to be found and their period derivatives need to be measured with accuracy \citep{asc19}. Finding such an object would lead to a wealth of studies to understand its formation and characteristics, e.g., testing the evolutionary scenarios of black holes and testing the relation between the stellar velocity dispersion and the mass of the black hole in a new range of mass values. The discovery of an MSP orbiting an IMBH would also give a unique chance to measure the black hole spin directly \citep[][see also Liu 2012, PhD, U. of Manchester]{Liu2014}.

If an IMBH forms a binary with either a stellar-mass black hole, a neutron star, or a white dwarf as a companion, the system emits gravitational waves in the LISA frequency range \citep{mhmjv2010, fgk2018, aac2021, sfb2023}. Hence, if discovered, an IMBH-pulsar binary would be a very promising source of gravitational waves in the LISA band, and targeted searches can be undertaken.

\item {\it Proper Motion}: Proper motions of almost all GCs have been measured using GAIA-DR2 \citep{bhs2019, vasliev2019}. Pulsar timing can provide alternate methods of measuring proper motions, and comparing measurements would be a valuable test of such measurements, as we elaborate below. 

The ensemble of pulsar proper motions, measured through pulsar timing, can be used to determine the cluster's proper motion and hence infer its orbit in the Galactic gravitational potential. While radio can provide accurate proper motion, optical instruments (especially 30-meter-class telescopes) can provide accurate radial motion. The combination would produce a 3D velocity vector for the GC. When this is done for a sizable number of GCs, it will be possible to constrain the Galaxy's gravitational potential well \citep{fw1980, hzt2015, bsb2021, dzo2026}. For nearby clusters, there is also the chance to detect the peculiar motions of the MSPs within each cluster. If these pulsars are in binary systems with optically detectable companions, then the optical radial motion and hence the 3D motion of the population of binary MSPs in a GC can be determined. Again, this will be a handle for constraining the GC potential well, especially in the core, where the relatively massive MSPs (compared to the mean stellar mass) typically reside. 

\item {\it Intra-cluster Medium}: GCs are expected to have a significant amount of ionized intra-cluster medium due to stripping off from old stars by stellar winds. However, observationally GCs contain very little gas that can not be fully explained by the ram pressure stripping process during their passage through the Galactic disk once in every $10^8 - 10^9$ yrs. Hence, it has been hypothesised that this medium is ejected out by the winds of neutron stars \citep{spergel1991, mb2011, NSfRr2020}. However, more numerical and observational evidence is needed to establish this scenario beyond doubt. 

Differential dispersion and rotation measures (DMs and RMs) between the detected pulsars can be used as highly sensitive, unique probes of this intra-cluster medium. These measures led to the first and the only detection of the ionised gas inside a GC \citep{Freire2001b,Abbate2018}, the potential discovery of turbulence in the gas and a tentative detection of the magnetic field \citep{Abbate2023}. This can teach us about the stellar winds of old stars releasing plasma in the cluster, the interactions between this plasma and the winds of pulsars and hot stars, leading to loss of the plasma from GCs and the growth of large-scale magnetic fields. Furthermore, the existence of this plasma, together with sensitive X-ray limits on the accretion luminosity from a central source, can be used to place an upper limit on the mass of a possible central black hole \citep{Abbate2018}. Thanks to the SKA telescopes, this accretion could eventually be detected directly in the radio band \citep{Karimi2024}. 

\item {\it Interaction History}: The types of pulsars that are found, e.g., their spin-period distribution, locations in the cluster, and the fraction of binary versus isolated pulsars, can vary quite drastically from cluster to cluster, and encode information about the dynamical history of the GC \citep{Verbunt2014}. Here we would learn about the still largely unknown stages of the evolution of a GC, with particular emphasis on the process leading to core-collapse \citep{Fregeau2008, Pooley2010}. It has been recently reported that pulsars in core-collapsed GCs rotate significantly slower than those in non-core collapsed GCs \citep{ohh2023}.

\item {\it Origin}: 
These systems will also allow the study of GCs not associated with our Galaxy. Nine of these GCs that are thought as a members of the Sagittarius Dwarf Spheroidal galaxy, namely NGC 6715, which is also the nucleus of this galaxy; Arp 2, Terzan 7, and Terzan 8, which are located in the main body of this galaxy; as well as Pal 12, Whiting 1, NGC 2419, NGC 4147, and NGC 5634, which are located in the extended tidal streams \citep{mrf2021}. None of these GCs presently has any known pulsars; however, these GCs may harbour a good number of MSPs as hinted by the $\gamma$-ray emission from the Sagittarius Dwarf Spheroidal itself \citep{cmm2022}. If discovered, pulsars in these GCs might help us probe the properties of those GCs and the Sagittarius dwarf spheroidal galaxy.

\end{itemize}

\subsection{Multiwavelength studies}

Deep multi-wavelength observations of GCs played a crucial role in pulsar science. For example, deep imaging of Terzan 5 \citep{Fruchter2000} spawned a great interest in that cluster, which later resulted in a record 49 known MSPs being found \citep{Ransom2005, Hessels2006, prf2024}.  Additionally, such multi-wavelength observations of GCs in X-rays \citep[primarily using Chandra; see e.g.,][]{Grindlay2001, Pooley2002, Grindlay2002, Heinke2003a, Heinke2003b, Heinke2005, Heinke2006, Bogdanov2006, Elsner2008, Bogdanov2011, zh22, lht23}, $\gamma$-rays \citep[primarily using Fermi; see e.g.,][]{Freire2011}, and optical \citep[primarily using HST; see e.g.,][]{Edmonds2001, Bassa2003, Bassa2004, Pallanca2010, Pallanca2013, Pallanca2014} provide complementary information on either the pulsar's magnetospheric emission, intra-binary emission (perhaps from a shock), or the companion itself. \citet{lht23} found that the relation between the X-ray luminosities with the spin-down energies is different for GC MSPs than that for the field MSPs. They also noticed an absence of any correlation between the spin period and the orbital periods of GC MSPs, while there is a strong correlation between these two parameters for the field MSPs. These two points hint at differences in the evolutionary past of MSPs in GCs and in the field. The next generation of extremely large telescopes may provide large numbers of radial velocities for constraining the mass ratios of the stellar components \citep{Cocozza2006}. Conversely, identifying high-energy GC sources as radio pulsars would help better understanding of the the zoo of objects that can be created (e.g., cataclysmic variables, low-mass X-ray binaries, etc.) and hence the stellar evolution history of the cluster itself \citep{Heggie2008}. For GC pulsars, we have an independent measure of the distance (and often the reddening), which can allow stronger constraints on various parameters of pulsars and their binary companions measured in optical, X-rays or $\gamma$-rays. The Fermi satellite is revealing $\gamma$-ray emission from several GCs \citep{Abdo2010, tkhcll11, dmcn19, wwxz22}. Comparing the total $\gamma$-ray emission with the number and properties of the known radio MSPs in the largest possible sample of GCs will be a new tool for investigating the as yet poorly understood high-energy emission mechanisms of the MSPs \citep[e.g.,][]{Harding2005}, as well as constraining the fraction of the sky swept by the MSP radio beam. Moreover, Fermi is not only detecting several GCs, but has started detecting individual pulsars, like PSR B1820$-$30A \citep{Freire2011}, PSR B1821$-$24A \citep{Johnson2013}, and PSR J1835-3259B \citep{zxw22}. We expect such multiwavelength studies to be continued in the era of the SKA telescopes.

\section{Observing Pulsars in GCs with the SKA telescopes}
\label{GCpulsarswithSKA}

\subsection{Searching for binaries: present status and future directions:}
\label{subsec:binarysearch}

Greater instantaneous sensitivity is not only important for finding weaker sources, but also because more than two-thirds of the known sources are in binary systems. Orbital motion smears the pulsar signal over multiple Fourier bins in the power spectra via the Doppler effect, and can make such sources undetectable unless this is corrected for \citep{jk1991}. For this reason, the intrinsic fraction of binary MSPs is likely larger than the observed fraction.

Typically, binary pulsars are discovered using an `acceleration search', which approximates the Doppler shift of the signal as a constant drift in the frequency domain \citep{Kulkarni1991, Ransom2001b}. This approach is relatively computationally efficient, whereas full searches of all Keplerian orbital parameters (even in the case of a circular orbit, where only three orbital parameters are needed) are currently not tractable for general use. Acceleration searches are only valid when the integration time of the full observation is less than about 
$10\%$ of the orbital period.  In other words, one cannot necessarily gain sensitivity to binary pulsar systems simply by integrating for a longer time \citep[though some analytical techniques do exist; see][]{Ransom2001b}. The shortest known orbital period of a GC pulsar is 53 minutes \citep{plj23}. Such a system can be found in a linear acceleration search of a 10-minute data set, only if the pulsar is bright enough to be detectable with the partial flux recovery. Otherwise, an acceleration-jerk search would be necessary to search for binary pulsars in compact orbits by accounting for a linear change in acceleration during an observation \citep{Bagchi2013}. One can consider an approximate 10 minute integration time, accompanied by an acceleration-jerk search, to be a good strategy to search for short-period binaries. On the other hand, one can use 2 hours as an appropriate integration time to search for long-period binaries or isolated pulsars. 

Note that, the assumption of a constant acceleration is useful even for eccentric systems. In fact, with the constant-acceleration assumption, highly eccentric systems are more easily detectable at most orbital phases than low-eccentricity systems\footnote{See Madsen 2013, MSc, UBC; https://circle.ubc.ca/handle/2429/44897}. On the other hand, acceleration-jerk searches are extremely useful \citep{Bagchi2013} in detecting very short-period binaries, especially if they are of low eccentricity, and about 10 pulsars have been discovered using acceleration-jerk searches \citep{ar18, trrca21}. Although such searches are computationally costly, there are ongoing efforts to improve the implementation of this algorithm \citep{anda20}, and we expect that it will be much easier by the time SKA telescopes come online.

\subsection{Propagation effects}

As in all pulsar surveys, interstellar propagation effects can strongly limit the detectability, especially at the shortest spin periods, where residual smearing in time due to uncorrected dispersive or scattering delay can broaden the pulse in time to the point where it is no longer detectable. Many of the densest, most massive GCs known are located in the Galactic bulge. The line-of-sight DM is often large ($\gtrsim 200$\,pc\,cm$^{-3}$), as is the expected scattering. To mitigate these effects, one can observe at a higher radio frequency. Taking the competing effect of the typically steep spectra of pulsars \citep[$S_{\nu} \propto \nu^{-\alpha}$, where $1 < \alpha < 3$;][]{Maron2000, Bates2013} into account, it turns out that the $1400 - 2000$ MHz band is well suited for searching GCs with DMs greater than about $100$\,pc\,cm$^{-3}$. Roughly 100 of the GCs visible by the SKA telescopes have expected DM greater than $100$\,pc\,cm$^{-3}$ \citep[according to the NE2001 model of][]{Cordes2002}, and are excellent targets for SKA-MID, which provides the maximum instantaneous sensitivity in the $1400 - 2000$ MHz band. For the remaining about $60$ GCs that are visible by the SKA telescopes with lower expected DMs, SKA-LOW (and/or SKA-MID at 800 MHz in some cases) will present an exciting opportunity.

\subsection{New or rarely used techniques with potential of becoming game-changers}

In recent years, there has been significant progress in the techniques involved in pulsar searches. Some techniques have already been established to be superior to the pre-existing ones and some are still in the nascent stage. Below, we discuss some of the recent developments that have the potential to be improved further and be very valuable in finding pulsars in GCs with data from the SKA telescopes.

\begin{enumerate}

\item {\em Imaging:} Imaging has already been successful in discovering pulsars in GCs \citep{urquhart2020, heywood2023, smirnov2024, smirnov2025, usc2026}. Most of these pulsars show variability, either due to eclipses \citep{smirnov2025} or refraction in the surrounding medium \citep{smirnov2024}. We expect this to be even more successful with the advanced imaging from the SKA telescopes.

\item {\em Analysis of baseband voltage data:} Algorithms to analyse high-resolution baseband voltage data from the MeerKAT and Effelsberg radio telescopes are being employed in a project whose primary aim is to discover two exotic classes of pulsars in GCs: (i) pulsars in highly relativistic, short-period orbits (a few hours or less) around compact objects and (ii) ultra-fast rotating pulsars with spin periods of about 1 ms or less. The project utilises the recently installed baseband voltage capturing extension at MeerKAT to record two-hour observations primarily targeted at core-collapsed GCs, which have a higher likelihood of hosting compact binary systems \citep{Verbunt2014}. Some of the innovations in the data analysis for this project include offline beamforming with various MeerKAT antenna subsets and across different Stokes parameters, enabling localisation of new pulsar discoveries and preserving full polarimetric information. Additionally, incoherent beam subtraction removes antenna autocorrelations, significantly mitigating radio frequency interference.

\item {\em More efficient searches for short-period binaries: } In addition to conventional acceleration and acceleration-jerk searches, effective for observations covering less than 10\% and 15\% of the orbital period, respectively, one can employ Keplerian-parameter searches to increase sensitivity when 15\% - 100 \% of the orbit is visible in an observation. For a circular orbit binary, this involves a three-dimensional search across the orbital period, projected semi-major axis and initial orbital phase. This technique was first used at the radio wavelengths by the Einstein@Home project to search for pulsars in the Pulsar Arecibo L-band Feed Array (PALFA) survey \citep{Knispel2013}. Recently, \citet{Balakrishnan2022} extended this algorithm to also be sensitive to elliptical orbit binaries, which includes two additional parameters, the longitude of periastron and the eccentricity. These Keplerian parameter searches utilise a stochastic template-bank algorithm, originally devised for gravitational-wave searches \citep{Messenger2009}, enabling optimal sampling and sensitivity across the entire orbital parameter phase space. These techniques are currently being implemented and advances in efficient gridding techniques will further optimise computational resources, ensuring these methods remain practical.

\item {\em Real-time folding versus offline beam-forming: } Real-time pulsar folding techniques enable the detection of periodic signals during data acquisition, minimising latency and improving response times for follow-up observations. Offline beam-forming, on the other hand, leverages stored search-mode data for post-acquisition analyses. This approach enhances sensitivity to weak signals, allowing astronomers to reprocess data with improved algorithms or under new hypotheses. 

\item {\em Application of distributed computing: } Volunteer-based distributed computing projects, such as Einstein@Home, have also played a significant role in pulsar discovery \citep{Knispel2013}. By harnessing the idle computing power of thousands of volunteers, these initiatives have shown that they can effectively be used to detect elusive binary pulsars in large datasets, as expected from the SKA telescopes.

 \item {\em Fast Folding Algorithms: } The detection of slow-rotating pulsars in GCs has been hindered in the past by excess red-noise in the long scans typically used, and also by inefficiencies in the Fourier-based search methods related to harmonic summing for narrow duty-cycle pulsations. However, various red-noise removal techniques are now routinely employed, and sensitivity to narrow duty-cycle pulsations is much improved due to efficient implementations of the Fast Folding Algorithm (FFA), such as \texttt{riptide} \citep{mbs2020}. We will likely see the detection of several more slow pulsars, including some in GCs, over the next several years as we apply these techniques to both archival data, as well as to the sensitive new data from SKAO.
    
In addition, recently, a number of very slow pulsar like objects, sometimes known as `long-period radio transients (LPTs)' have been discovered \citep[and references therein]{wus25, hrm23}. It is still not well established whether these sources are neutron stars or white dwarfs. So far, all discovered LPTs (7 as of March 2025, see Fig 4 of \citet{wus25}) are in the Galactic field. It would be interesting to find such sources in GCs, and regular use of single pulse searching and the FFA might allow that to happen, especially in combination with the increased sensitivity of the SKA telescopes.

\end{enumerate}

\subsection{Array Configuration}

GC pulsar searches will require a tied-array observing mode of the SKA telescopes because the data must be recorded with at least $\sim 50$\,$\mu$s resolution. Maximum possible sensitivity can be achieved if the data are coherently dedispersed at the average cluster DM (in the case of GCs with known pulsars) or at least at a few trial DMs for clusters with no known sources. In practice, this will likely require VLBI-like ability to record raw voltages from the tied-array. An important consideration is the sky area that must be covered for such searches. The known Galactic GCs typically have core radii of a few arc-seconds to 2 arc-minutes and half-light radii of about 0.5 to 3 arc-minutes \citep{Harris1996}. GC pulsars congregate to a large extent towards the cluster core, and almost all of the known GC pulsars are within one arc-minute of the cluster's optical centre-of-light. The 1000 m (radius) core of SKA-MID AA$^*$ will provide a tied array beam with FWHM of $\sim 0.4$ arc-minute at 1400 MHz, meaning that 16 tied-array beams (as allowed by the PST back-end) for each observation is enough to detect and discover most of the GC pulsars using this setup.  For the 600-m (radius) core of SKA-LOW, a single $\sim 3$ arc-minute tied-array beam at 350 MHz will be sufficient. In order to catch the minority of pulsars located further from the core, which are still very interesting for understanding the dynamical history of the cluster, a mosaic of pointings for any given cluster will be needed at 1400 MHz. In summary, searching for GC pulsars is an excellent early science and commissioning project, which will naturally pave the way to an all-sky pulsar survey, which will require the collection of data from several hundreds of thousands of tied-array beams. 

More ambitious searches using a larger fraction of the SKA-MID/LOW antennae will be more challenging, although not unfeasible for a subsample of the GCs. For example, an almost 50\% improvement in the instantaneous sensitivity provided by SKA-MID (in both the AA$^*$ and AA4 configurations), would require going from a 1000 m (radius) core to a radius of about 8000 m from the array centre, and thus $64$ times more beams to cover the same area of sky. The total number of required beams is thus still compatible with the investigation with a single pointing of the area of the core, only for the sample of the most compact core collapsed GCs. 

A full-array configuration is certainly desirable for the follow-up of new pulsars that have already been sufficiently well localised. In fact, the repeated timing observations that are needed to extract the science (e.g., precision astrometry, determination of the Keplerian and post-Keplerian orbital elements, measurement of the proper motion, etc.) can be done by using one narrow, full-array tied-array beam on each source.

\subsection{Achievable sensitivity}

The 1000-m (radius) core of the SKA-MID AA$^*$, operating from $1250 - 1550$ MHz, has a SEFD of about 5.3 Jy, obtained by scaling from  \citet{bailes2020meerkatfacility}. Thus, it can achieve an rms noise of about 2.6 $\mu$Jy for a 2-hr integration. Using a digital beam-former and back-end similar to those sketched in the SKA Baseline Design (i.e., capable of keeping the smearing effects of interstellar dispersion smaller than the adopted sampling time of 50 $\mu$s for all DMs less than $1000$\,pc\,cm$^{-3}$), that rms noise translates into a limiting sensitivity (at ${\rm S/N} = 8$) of about $8$ $\mu$Jy for a recycled pulsar spinning at about 1 ms period and having an intrinsic pulse duty cycle of $10\%$.  For the case of the SKA-MID AA4 configuration (SEFD of about 3.9 Jy for the 1 km (radius) core), the corresponding limiting sensitivity is about 6 $\mu$Jy. Analogously, using the full AA$^*$ array, the limiting sensitivity will be about $5$ $\mu$Jy, whereas for the full AA4 array, the limiting sensitivity will be about 3.5 $\mu$Jy. Sensitivity limits would be halved from those quoted above for the GCs where 8-hr long tracks with the SKA-MID are possible (this, fortunately, includes some of the most promising targets).  

The 600-m (radius) core of the SKA-LOW, operating from $250-450$\,MHz, can achieve about $30$\,$\mu$Jy sensitivity for 2 hr integrations (depending on the line-of-sight; for clusters in the Galactic bulge, the sky temperature will reduce the sensitivity a bit). For comparison, the all-sky GBNCC survey at 350\,MHz achieved about $1$\,mJy sensitivity \citep{Stovall2014}.  

\section{Available discovery space}
\label{subsec:discoveryspace}

The SKA telescopes will provide unprecedented sensitivity for targeted GC searches in the southern skies, which is fortunately also where the majority of the most massive, densest, and hence most pulsar-rich GCs can be found, i.e., in the Galactic bulge. While Galactic field pulsars are spread across the full 41,000 sq. degrees of the celestial sphere, the total area required to search all known Galactic GCs combined constitutes only about 1 sq. degree!  Hence, compared with the greater observing and processing challenges of an all-sky search, which will finally require the analysis of several hundreds of thousands of tied-array beams, deep targeted searches of these limited fields-of-view can be handled with only about thousands of beams. 

MSPs are in general weak radio sources (phase-averaged flux densities at 1400 MHz, i.e., $S_{1400}$, being less than 1 mJy), and GCs are typically at distances larger than $4$\,kpc. Raw sensitivity and long dwell time is thus crucial for finding the weakest cluster pulsars. The weakest known GC pulsars have $S_{1400}  \lesssim 10$\,$\mu$Jy and were discovered in multi-hour integrations using the Arecibo, GBT, MeerKAT and FAST radio telescopes with several hundred MHz of bandwidth at 1400 or 2000 MHz.  This sets the bar that the SKA telescopes must exceed, implying that the integration times will need to be significantly longer than for the planned all-sky searches: i.e., some hours versus tens of minutes. This in turn requires that the data will have to be collected in the baseband mode by the PST back-end, which will be able to provide up to 16 simultaneous beams for SKA-MID\footnote{For SKA-LOW only 8 beams will be available in AA$^*$ configuration, later raising to 16 for AA4 configuration.}. The 16 streams of data will then be promptly transferred to the network of SKAO Regional Centres and analysed with the pipelines installed there. These pipelines are already available and well tested on the ongoing experiments. In addition to ensuring effective commissioning of the data collection from the SKA telescopes, this will provide early science by quickly finding at least dozens with AA$^*$ configuration, and eventually a few hundred pulsars in GCs with AA$4$ configuration. The expected scientific outcome is very promising, leading to the publication of up to a couple of hundred pulsar discoveries in the first months of operation of the telescope.  

The transfer of the data to the Regional Center will allow to keep them for subsequent searches with improved codes as well as to re-use them to time new pulsars which will be discovered in each of the clusters when a new delivery stage of the SKA configurations will be made available, in particular, it is foreseen that a survey with AA4 will determine a new blooming of discoveries with respect to the first burst of new pulsars resulting from the observations with AA$^*$.

\subsection{Prediction based on the established correlations with GC properties}
\label{subsec:GCpropertiesPredictions}

We know that we are currently only sampling the tip of the pulsar luminosity distribution in these predominantly distant ($\gtrsim 4$\,kpc) stellar systems, i.e., GCs. In fact, various investigations \citep{Bagchi2011, Chennamangalam2013,martsen22} confirmed that the luminosities (defined as $L_{\nu}=S_{\nu} \, d^2$, where $S_{\nu}$ is the observed flux density at a central frequency ${\nu}$, and $d$ is the GC distance) of the pulsars observed in a GC can be reproduced as the bright tail of either a log-normal distribution, with parameters compatible with the luminosity functions (LFs) inferred for the pulsars in the Galactic field, or a power-law distribution with the index of about $-1$ \citep[in agreement with earlier results by][]{McConnell2004, Hessels2007}, with the former functional form providing a slightly better match to the available data.  For both the assumed LFs, in a large range of not too weak luminosities (typically above 0.5 mJy kpc$^2$), the flux density distribution follows ${\rm d} \log N/{\rm d} \log S_{\nu} \sim 0.5-1$ for any given cluster. That implies that a large increase in the sensitivity will automatically bring many new sources within the reach of detection and will also provide higher-precision studies of the (mostly very faint, $S_{\rm 1400} \sim 20$\,$\mu$Jy) sources that are currently known.  This will also enable more mass measurements and equation-of-state constraints using already known sources. 

As mentioned in Section \ref{subsec:population_clusterproperties}, to quantify the number of pulsars expected in a search of all Galactic GCs, using the SKA telescopes, we used the scaling relationship $N = \Gamma^{0.7}$ \citep{hct2010} to first estimate the total number of pulsars across all known systems: the implied number is roughly $6000$. This is more optimistic than the predictions based on detailed population synthesis simulations \citep{Turk2013}, which suggest a conservative total of $600 - 3700$ detectable (i.e., beamed towards us) pulsars in the Galactic GCs. However, in this section, we use the aforementioned scaling to get an estimate based on the least possible number of underlying assumptions.  Moreover, the proximity of the numerical values of the escape velocities and two-body encounter rate ensures the fact that a replacement of the above relation with a relation between the pulsar content and escape velocity (as reported by \citet{Yin2024}) would not change the results much. In a subsequent section, we will present a more conservative approach, providing a closer match to the predictions of \citet{Turk2013}. 

\begin{figure}[h]
    \centering
	\includegraphics[width=0.5\columnwidth]{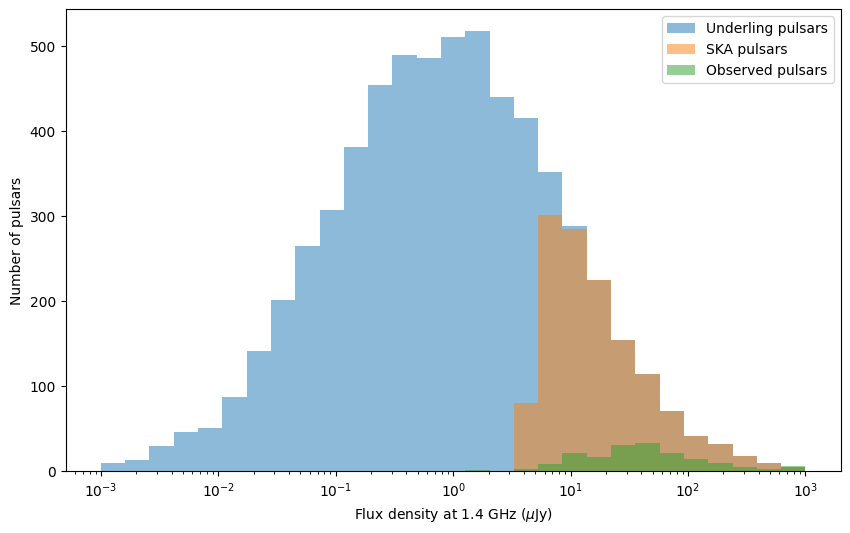}
    \caption{Simulated GC pulsars detectable in a 2~hr baseline SKA-MID (AA4 configuration) survey shown alongside the underlying model population and the current sample of GC pulsars with measured flux densities. Although a number of faint pulsars will still remain undetectable with such a survey, the enhancement in the size of the population of high flux density pulsars is significant.}
    \label{fig:GCsim}
\end{figure}

With the sample obtained using \citet{hct2010} model, we assume a 2-hour integration time survey of all clusters with 350~MHz of bandwidth and 5.8~K/Jy sensitivity (i.e., AA4 specifications). In Fig. \ref{fig:GCsim}, we show the histogram of potentially detectable pulsars in all GCs. This analysis assumes a log-normal L-band luminosity function described by \citet{Faucher2006} and results in a total of around 1700 detectable pulsars. This represents about 5 fold increase in the sample
size at present. As can be seen in the results of this analysis in Fig.~\ref{fig:GCsim}, a significant fraction (about 4/5) of the predicted population of GC pulsars are below the thresholds of the baseline sensitivity. Deeper observations, for example, 8-hour tracks, of the closest 10 GCs in the sky are estimated to probe these clusters much more completely. Using the above assumptions, we estimate that around 2/3 of the 300 potentially observable pulsars in these GCs would be detectable. In some cases, it might be possible to detect all the active radio pulsars in a given cluster, providing a unique view of the star formation history and interactions over the cluster's lifetime. 

Note that, about $98\%$ of the known Galactic GCs are visible to SKA-MID/SKA-LOW for at least 2 hours, assuming elevation limits of 15/30 degrees, respectively. For 8-hour integrations, there are still 141/113 clusters available for observations at SKA-MID/SKA-LOW, respectively. Of these, 45 nowadays have at least one known pulsar, and hence the DM of the cluster is also known (a fact that makes searching for pulsars significantly easier). 

\subsection{Predictions based on improvements of telescope sensitivities}
\label{subsec:conservativeprediction}

\begin{figure}[h]
    \centering
	\includegraphics[width=0.48\columnwidth]{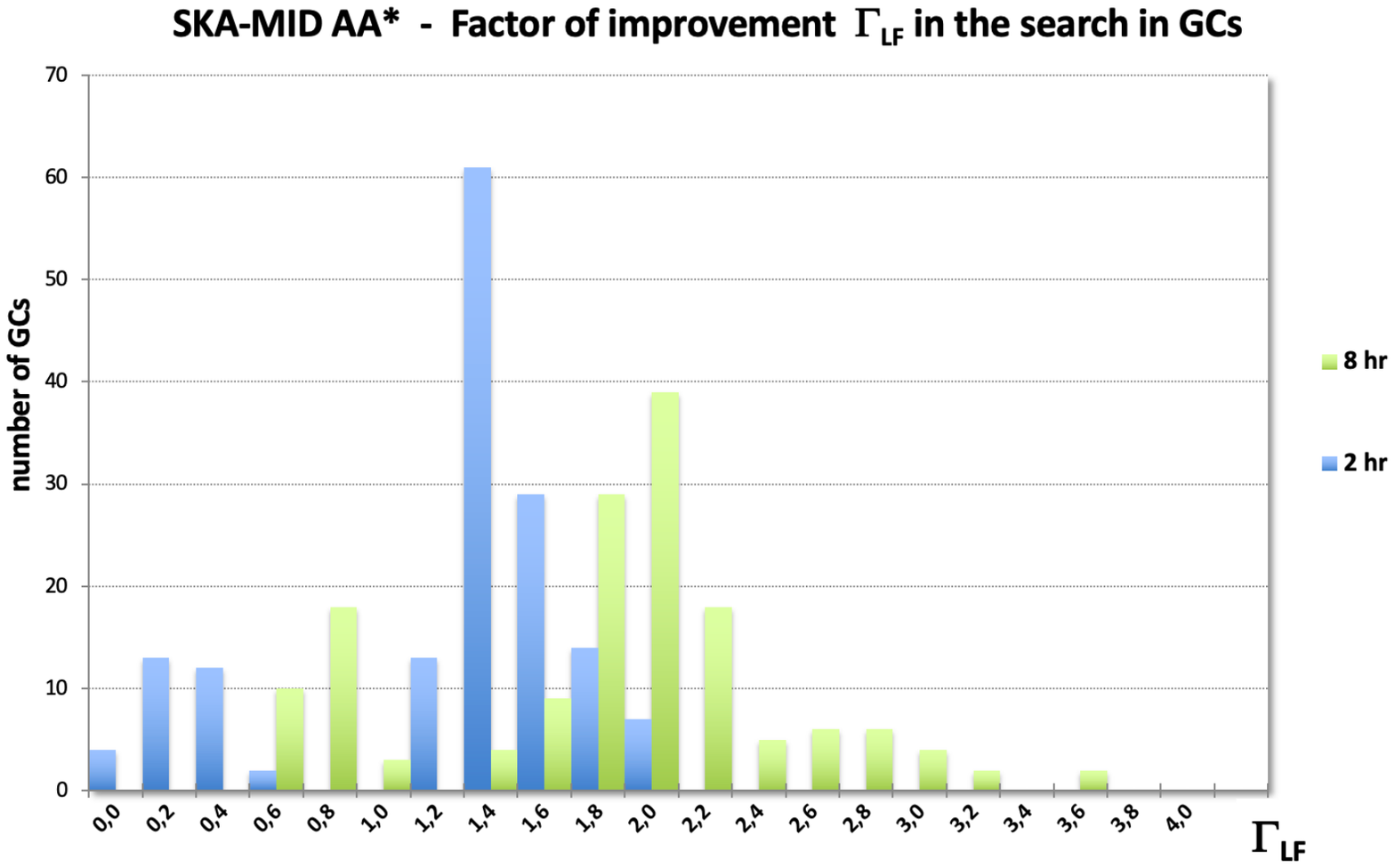}
    \hfill
	\includegraphics[width=0.48\columnwidth]{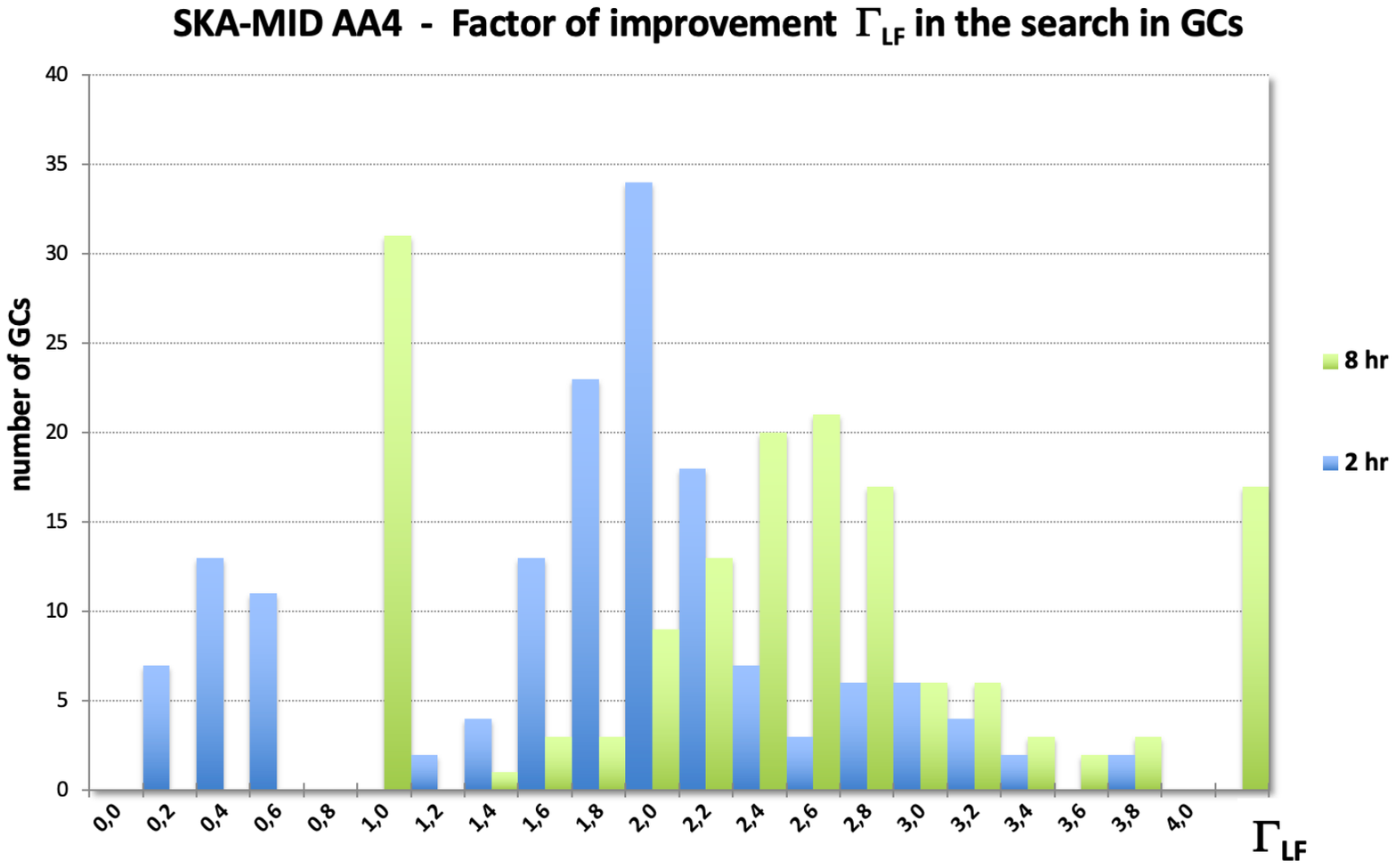}
    \caption{Left: distribution of $\Gamma_{LF}$, which represents the growth of the probed area of the pulsar luminosity function, for all GCs visible to the SKA telescopes. Here we assume the use of SKA-MID AA$^*$ in 2-hour (blue) and 8-hour (green) observations, compared to the state-of-the-art observations with FAST and MeerKAT. Right: distribution of $\Gamma_{LF}$ assuming the use of SKA-MID AA4 in 2-hr and 8-hr observations, compared to the state-of-the-art observations with FAST and MeerKAT. See text for further details.}
    \label{fig:SKA_AAstar_improvement}
\end{figure}

Fig. \ref{fig:SKA_AAstar_improvement} illustrates the discovery space that is open to SKA-MID for pulsar searches in GCs (a similar analysis can be made for SKA-LOW, too). We define the quantity $\Gamma_{LF}$ to represent the growth of the probed area of the pulsar luminosity function (LF) of a given GC. This assumes a survey performed with SKA-MID at the central frequency of 1400\,MHz, with 300\,MHz bandwidth and other survey parameters like those in the SKA baseline designs. In particular, the histogram of Fig. \ref{fig:SKA_AAstar_improvement} is obtained by comparing the sensitivity of the SKA-MID survey with that of the best possible GC searches already carried out or still ongoing, especially, for each cluster, we assume that the best-possible search can be conducted by either FAST or MeerKAT, depending on the declination of the cluster (FAST for the GCs in the Northern sky and MeerKAT for the GCs in the Southern sky). All flux density limits of the surveys have been scaled to 1400\,MHz using a pulsar spectral index $-1.7$.  We also assumed that the effects of the dispersion and scattering are the same in all instruments/surveys. 
Using these assumptions, and an effective pulsar duty cycle of 25\% (which is typical for millisecond pulsars), we calculated survey flux density limits. The next step in producing Fig. \ref{fig:SKA_AAstar_improvement} was to feed the calculated sensitivity limits to a log-normal pulsar LF with the mean (in units of mJy\,kpc$^2$ expressed in a logarithmic base-10 scale) of $-1.1$ and the standard deviation 0.9, which is known to reproduce the observed data \citep{Bagchi2011,Chennamangalam2013}.

We considered the distributions of $\Gamma_{LF}$ for 2-hour and 8-hour integrations, colour-coded in blue and green, respectively.  The average factor $\Gamma_{LF}$ over the entire population is between about $1.4$ and $1.7$ (see also Table \ref{tab:predictionNumbers}), thus suggesting that a survey over the GCs with SKA-MID would provide a 40\%-70\% improvement in probing the LF (the only exception for some GCs inspected by FAST). With all the caveats related to the diversity of the features shown by the pulsar population hosted in GCs (each feature having its own impact on the discovery biases), one can extrapolate about 150 or 250 discoveries in a nearly 2-hour survey with SKA-MID AA$^*$ or AA$4$ looking at the 45 GCs which are already known to contain pulsars. This is an easy-to-implement program for the initial phase of the SKA-MID AA$^*$ activities, in fact, it needs a very small number of beams, plus a modest amount of the telescope time, plus a limited request of the computational power in the SKAO Regional Center network, since the dispersion measures of the aimed GCs are already known (thus drastically reducing the computational time).  We note that, with 8-hour integrations, SKA-MID AA4 can much more deeply probe $(\Gamma_{LF}>4)$ a bunch of very distant GCs (typically at several tens of kpc from the Earth), the luminosity function of which can only be marginally probed by the current experiments.

\begin{table}[]
    \centering
 \begin{tabular}{ |c|c|c| } 
 \hline  \hline
\multirow{ 4}{*}{Configuration}  & Average improvements  &  \\  
  &  in the explored area & Total discoveries \\  
  &  of the GC pulsar & (in GCs with  \\ 
  &  luminosity factor &   already known pulsars) \\   \hline
 SKA-MID AA* core 2 hr & 0.95 & $\sim 20$ \\ 
  SKA-MID AA* core 8 hr & 1.3 & $\sim 100$ \\ 
  SKA-MID AA* 2 hr & 1.4 & $\sim 150$  \\ 
  SKA-MID AA* 8 hr & 1.7 & $\sim 200$ \\ 
 SKA-MID AA4 core 2 hr & 1.2 & $\sim 100$  \\ 
 SKA-MID AA4 core 8 hr & 1.7 & $\sim 200$  \\ 
SKA-MID AA4 2 hr & 1.9 & $\sim 250$ \\ 
 SKA-MID AA4 8 hr & 2.4 & $\sim 300$ \\ \hline
 \hline
\end{tabular}  
    \caption{Expected number of discoveries of additional pulsars in GCs which are already known to host pulsars, for various configurations of SKA-MID (at $1400$\,MHz), using 2-hour and 8-hour observations. The configurations labelled ``core'' refer to the use of only the 1 km radius central part of the array. }
    \label{tab:predictionNumbers}
\end{table}

In contrast with the calculation performed in Sec \ref{subsec:GCpropertiesPredictions}, we note that, by using the approach adopted in this section, a direct extrapolation from $\Gamma_{LF}$ to the expected total number of pulsar discoveries in the whole population of GCs is impossible, since an estimate cannot be obtained for the GCs where no pulsar is known so far (i.e., the vast majority of objects in the Galactic GCs). Furthermore, use of a power-law luminosity function (in place of the adopted log-normal luminosity function) results in more than doubling all the figures in Table \ref{tab:predictionNumbers}. In view of all the considerations above, we finally remark that the reported $\Gamma_{LF}$ can only be regarded as a conservative proxy for comparing the capabilities of SKA-MID in the search for GC pulsars with respect to the capabilities of the current state-of-the-art experiments/telescopes, i.e., MeerKAT and FAST. Notwithstanding, the number of discoveries resulting from the direct multiplication of $\Gamma_{LF}$ with the number of already detected objects is in the range of the predictions of much more sophisticated analyses \citep[e.g.,][]{Bagchi2011, Chennamangalam2013} developed to predict the expected pulsar yield in some of the most populated GCs. 

\section{The Importance of Archival Search-Mode Data}

Given that we are nearly guaranteed that the central parts of GCs contain dozens to even hundreds of detectable but still undetected pulsars, even in the era of the SKA telescopes, it is important to consider special handling of at least some of the tied-array beams from these observations. We strongly encourage SKAO to enable long-term archiving of at least the central beams of all GC observations.

New algorithms and improved computing enable new pulsars to be discovered in archival GC observations \citep[e.g.,][]{ar18,crf18}. Yet even more importantly, when new pulsars are discovered in a cluster, once their spin frequencies and DMs are known, one can often uncover those new pulsars in archival observations in order to solve orbits, look for orbital variations, enable ``instantaneous'' long-term timing \citep[e.g.,][]{frk2017,ridolfi2022ngc1851disc,vsb22,dpr22,prf2024}, etc. This process works even if the signal-to-noise ratios of the archival detections are as small as about 5. These opportunities do not exist for wide-area pulsar surveys, where the chances of a single phased-array beam containing multiple pulsars is small. While the amount of data that needs to be saved is not insignificant, the central beams can be prioritised where the chances that they contain as-yet-undetected new pulsars are the greatest.

To estimate the nominal amount of data storage needed, we assume that we will perform a survey of all 157 known GCs with 2 epochs per cluster, 16 beams, 4096 frequency channels and  7200 seconds of observations. Assuming a nominal sampling time of $75 \mu$ seconds, this results in a total of 2 PB of data. Since some of these clusters already have known pulsars, and we will likely be able quickly to find the first pulsars in many others, we assume that 70\% of this data can be sub-banded at the DM of the cluster to about 256 channels on average. This will result in the total long-term storage of around 700 TB. 

Apart from survey observations, the storage of sub-banded data obtained during regular timing or follow-up observations is also extremely crucial for performing deeper searches of the clusters and for confirming any future discoveries. Assuming that we observe on average 100 GCs for 5 epochs a year, 7200 seconds per observation, a sampling time of 75~$\mu$s, 16 PST beams and 256 sub-banded channels, we obtain a total data size of 1 PB/year. While this data rate is preferred, we can reduce the number of beams and further sub-band the data to obtain a data rate of about 250 TB/year. 

\section{Conclusions}

Once construction is completed, SKA-MID and SKA-LOW will be the premier search machines for pulsars in GCs. Their raw sensitivity will surpass that of MeerKAT and even FAST, with the additional advantage that most of the GCs will be visible by the SKA telescopes (unlike FAST). As we have seen in Section \ref{subsec:conservativeprediction}, even a conservative approach predicts discoveries even only with the core of SKA-MID AA*, and the full AA* or eventually AA4 is expected to increase the number of discoveries even more. This offers a great opportunity for early SKAO pulsar science, even before all the collecting area is in place. At the same time, a more optimistic prediction calls for up to 1700 pulsars to be detectable with SKA-MID AA4 configuration in all Galactic GCs visible by the SKA telescopes.

The predicted significant enhancement of the GC pulsar population will translate in finding more and more exceptional pulsar systems, in turn promoting new tests of strong gravity theories, dense matter, and fundamental physics. For example, there are 10 GC pulsars out of a total 345 having spin frequency larger than 500 Hz (the number of such ultra-fast pulsars being 35 out of total 3781 known pulsars). Interestingly, 8 out of 10 ultra-fast GC pulsars are located in the globular cluster Terzan 5! An enhancement in the population of GC pulsars thus has the prospect for breaking the current 716-Hz rotation record held by Ter5ad. Besides the study of individual pulsar `jewels', the ensemble of detected pulsars will also provide a unique probe to the dynamics, evolution, gas content, and the magnetic field configurations of the galactic GCs.

\section*{Acknowledgements}

WWZ is supported by the National SKA Program of China No. 2020SKA0120200 and the National Nature Science Foundation of China (grant No. 12041303). FA, MCiB \& AP acknowledge that part of this work has been funded using resources from the INAF Large Grant 2022 {\it GCjewels} (PI Andrea Possenti) approved with Presidential Decree 30/2022. AP also acknowledges that this work was supported in part by the `Italian Ministry of Foreign Affairs and International Cooperation' grant number ZA23GR03, under the project {\it RADIOMAP-Science and technology pathways to MeerKAT+: the Italian and South African synergy.} FA acknowledges that part of the research activities described in this paper were carried out with the contribution of the NextGenerationEU funds within the National Recovery and Resilience Plan (PNRR), Mission 4 – Education and Research, Component 2 – From Research to Business (M4C2), Investment Line 3.1 – Strengthening and creation of Research Infrastructures, Project IR0000034 – `STILES -Strengthening the Italian Leadership in ELT and SKA'.

\bibliographystyle{abbrvnat-maxbibnames4}
\bibliography{GCpsrSKA2024}

\end{document}